\documentclass[doublecol]{epl2} 

\date {}
\usepackage{amsfonts,amsmath,epsfig}
\newcommand{\be}{\begin{eqnarray}}
\newcommand{\ee}{\end{eqnarray}}

\def\kk{{\bf k}}

\title{Electron pair forming in an integrable model}
\author{Jan Naudts \and
Tobias Verhulst\footnote{Research Assistant of the Research Foundation - Flanders (FWO - Vlaanderen)}
}
\institute{
Departement Fysica, Universiteit Antwerpen,
Groenenborgerlaan 171, 2020 Antwerpen, Belgium}


\abstract{
We consider a lattice model in which phonons scatter with pairs of electrons.
All eigenvalues and eigenvectors can be obtained analytically. For a suitable
choice of parameters the ground state consists of a Fermi sea of non-interacting
electrons, with on top a layer of paired electrons. The binding energy
of the electron pairs is partly cancelled by an increased kinetic energy.
This results in a gap in the kinetic energy spectrum of the electrons.
}

\begin{document}
\maketitle

\section{Introduction}
It was recently suggested \cite {PC09,PCC09} that the binding energy of Cooper pairs is not constant.
Breaking one Cooper pair has an effect on all other Cooper pairs, due to the
Pauli blocking of the composite bosons. In the present note we introduce a simple microscopic
model in which the binding energy of the paired electrons is partly cancelled by an increasing
kinetic energy, in such a way that the effective binding energy is indeed not constant.

The sole interaction term of the model is of fourth order in the electron operators and
quadratic in the phonon operators. This means a scattering between phonons and electron pairs.
One would expect such a term in the context of the bipolaronic Hamiltonian \cite {AR81,AD09}.
But here, the term is responsible for the formation of electron pairs, rather than modelling interactions
between existing pairs.

\section{The model}
The motion of the electrons is described by the hopping term. In $\kk$-space
it can be written as
\be
H_{\rm el}=-\sum_{\kk}\hbar\lambda_\kk \left(B_{\kk,\uparrow}^\dagger B_{\kk,\uparrow}+B_{\kk,\downarrow}^\dagger B_{\kk,\downarrow}\right).
\ee
The coefficients $\lambda_\kk$ are real and satisfy $\lambda_{-\kk}=\lambda_\kk$
and $\sum_\kk \lambda_\kk=0$.
The lattice phonons are also needed with two different polarisations. For the sake
of convenience they are denoted with up and down arrows. Their contribution to the Hamiltonian is
\be
H_{\rm ph}=\sum_{\kk}\hbar\mu_\kk\left(c_{\kk,\uparrow}^\dagger c_{\kk,\uparrow}+c_{\kk,\downarrow}^\dagger c_{\kk,\downarrow}\right).
\ee
The operators $c_{\kk,\uparrow}^\dagger,c_{\kk,\uparrow},c_{\kk,\downarrow}^\dagger, c_{\kk,\downarrow}$ satisfy the canonical
commutation relations.

Introduce pseudospin operators
\be
J_\kk^{-}
&=&B_{-\kk,\uparrow}^\dagger B_{\kk,\downarrow}^\dagger B_{-\kk,\downarrow} B_{\kk,\uparrow}\crcr
J_\kk^{+}
&=&B_{\kk,\uparrow}^\dagger B_{-\kk,\downarrow}^\dagger B_{\kk,\downarrow} B_{-\kk,\uparrow}\crcr
J_\kk^z&\equiv&\frac 12[J_\kk^{+},J_\kk^{-}]\cr
&=&\frac 12n_{\kk,\uparrow}(1-n_{-\kk,\uparrow})n_{-\kk,\downarrow}(1-n_{\kk,\downarrow})\crcr
& &
-\frac 12n_{\kk,\downarrow}(1-n_{-\kk,\downarrow})n_{-\kk,\uparrow}(1-n_{\kk,\uparrow})
\ee
with $n_{\kk,\tau}=B^\dagger_{\kk,\tau}B_{\kk,\tau}$.
They satisfy the su(2) relations
\be
[J_\kk^z,J_\kk^\pm]=\pm J_\kk^\pm.
\ee
The operator $J_{\kk}^{+}$ flips a spin current from
the direction $-\kk$ into the direction $\kk$.

Introduce also new bosonic operators by
\be
a_\kk
=c_{\kk,\uparrow}c_{-\kk,\downarrow}^\dagger,
\quad\mbox{ and }\quad
a_\kk^\dagger
=c_{-\kk,\downarrow}c_{\kk,\uparrow}^\dagger.
\ee
Some care is needed when doing calculations with these operators because they do not satisfy the
canonical commutation relations.

The model Hamiltonian is now
\be
H=H_{\rm el}+H_{\rm ph}+\hbar\xi\sum_\kk \left(a_\kk^\dagger J_\kk^-+a_\kk J_\kk^+\right).
\label {modham}
\ee
Other terms can be added without spoiling the integrability
of the model. However, they are not needed for understanding the model.

\section{Eigenstates}
The interaction term of (\ref {modham}) involves only electrons and phon\-ons
with the same or opposite momentum. Hence, the determination of eigenstates reduces
to an easy problem involving at most 4 electrons.
The electronic subsystem with
momenta $\pm\kk$ has 16 possible states, 14 of which do not interact with the phonons.
The only two states that do interact are two-electron states with total momentum zero and
total spin zero. If the coupling constant $\xi$ is large enough then the ground state
of this two-particle subsystem involves a single phonon, which itself is a superposition of
two phonons with opposite momenta and polarisation. This ground state describes an electron
pair which is kept together by a deformation of the lattice.

\section{Ground state}
For the sake of simplicity the following argument is made for a one-dimensional model.
Let us start from the situation in which all states of the free electron model
are occupied up to the Fermi level. This is, $n_\kk=2$ if $|\kk|\le k_F$, and $n_\kk=0$ otherwise.
Introduce the notations $\lambda_F\equiv\lambda_{k_F}$ and $\mu_F\equiv\mu_{k_F}$.
Similarly, the derivatives are denoted $\lambda_F'\equiv\lambda_{k_F}'$ and $\mu_F'\equiv\mu_{k_F}'$.
In the following it is assumed that $\lambda'_F<0$.
Make the following expansions
\be
\lambda_k=\lambda_F+\lambda_F'(k-k_F)+O\left((k-k_F)^2\right),\cr
\mu_k=\mu_F+\mu_F'(k-k_F)+O\left((k-k_F)^2\right).
\ee

Take the 4 electrons with $\kk=\pm k_F$ and move two of them to the pair state at $\kk=\pm k_F$,
the other two to the pair state at $\kk=\pm(k_F+\delta k)$. Before moving, the 4 electrons together have
the energy $-4\lambda_F$. After moving, the pair at $\kk=\pm k_F$ has energy $\mu_F-2\lambda_F-\xi$.
The pair at $\kk=\pm(k_F+\delta k)$ has energy $\mu_F-2\lambda_F-\xi+(\mu_F'-2\lambda_F')\delta k$.
The gain in energy is
\be
2\xi-2\mu_F-(\mu_F'-2\lambda_F')\delta k.
\ee
Repeat the operation now removing 4 electrons from the free states at $\kk=\pm (k_F-\delta k)$
and making one pair with unchanged momenta, and one pair at $\kk=\pm (k_F+2\delta k)$.
Now, the gain in energy is only
\be
2\xi-2\mu_F-(\mu_F'-6\lambda_F')\delta k.
\ee
This procedure is repeated $n$ times, after which no more energy can be gained by creating electron pairs.
The condition for $n$ reads
\be
0=2\xi-2\mu_F-\left[\mu'_F-2(2n-1)\lambda'_F\right]\delta k.
\ee
The ground state has now been obtained. For $|\kk|<k_F-n\delta k$ all free electron states are
occupied. For $k_F-n\delta k <|\kk|<k_F+n\delta k$ the paired states are occupied.
The picture is not so different from that of a layer of Cooper pairs above a quiescent Fermi sea \cite {CLN56}.
The main difference is that in the present derivation the paired electrons do not all have the same energies. 
See the figure.

\begin{figure}
\onefigure[width=7cm]{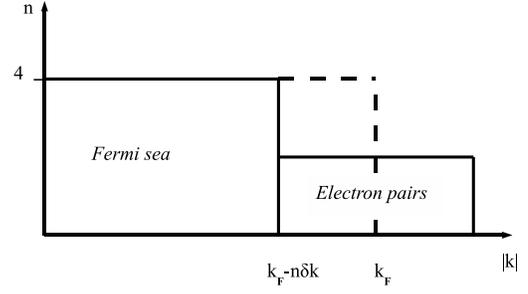}
\caption{Number of electrons as a function of the $k$-vector
in a one-dimensional ground state.}
\label {kspace}
\end{figure}

\section{Elementary excitations}

From the above construction of the ground state it is clear that the excitations, lowest in energy,
are the collision of two electron pairs, one at $\pm(k_F-(n-1)\delta k)$, the other at $\pm(k_F+n\delta k)$, and
the inverse process of 4 electrons at $\pm(k_F-n\delta k)$ forming two electron pairs. During these processes
the two electrons of one of the pairs lose or gain a momentum of approximately $2n\delta k$. This means
that there is a gap in the spectrum of the kinetic energy of the electrons.

\section{Discussion}

This note explores a novel mechanism for electron pair formation due to electron-phonon interaction.
The interaction term is unusual in that it is fourth order in the electron operators and
quadratic in the phonon operators. With a suitable choice of parameters the ground state exists
of a non-interacting Fermi sea with 4 electrons per pair of wave vectors $(\kk,-\kk)$,
and above this, a region with one electron pair per pair of wave vectors $(\kk,-\kk)$.
This results in a gap in the kinetic energy spectrum of the electrons, but not in the spectrum of
the total Hamiltonian.

More details about the present model and its extensions will be published elsewhere.


\end{document}